\documentclass[conference]{IEEEtran}
\IEEEoverridecommandlockouts
\usepackage{cite}
\usepackage{amsmath,amssymb,amsfonts}
\usepackage{algorithmic}
\usepackage{graphicx}
\usepackage{textcomp}
\usepackage{xcolor}
\def\BibTeX{{\rm B\kern-.05em{\sc i\kern-.025em b}\kern-.08em
    T\kern-.1667em\lower.7ex\hbox{E}\kern-.125emX}}
\usepackage{amsfonts,algorithm,algorithmic,amsmath,amsthm,amssymb,graphicx,algorithm,epstopdf,cite,url,subfigure}
\usepackage{psfrag,amssymb,amsmath,pifont,cite,graphics,graphicx,epsfig,subfigure,amscd,helvet,multirow,enumerate}
\usepackage{lmodern}
\usepackage{bm}
\usepackage{color}
\usepackage{epstopdf}
\usepackage{accents}
\usepackage{amssymb,amsmath,amscd,cite,enumerate}
\usepackage{graphics,graphicx,epsfig,psfrag,subfigure,threeparttable,url}
\usepackage[colorlinks,citecolor=black,linkcolor=black]{hyperref}
\usepackage[english]{babel}

\newtheorem{theorem}{Theorem}
\newtheorem{lemma}{Lemma}

\newtheorem{remark}{Remark}

\newcommand{\sgn}{\operatornamewithlimits{sgn}}

\newcommand{\supp}{\operatornamewithlimits{supp}}

\newcommand{\andd}{\operatornamewithlimits{and}}

\begin{document}

\title{On the Fundamental Recovery Limit of Orthogonal Least Squares
\thanks{This work was supported in part by the Samsung Research Funding \& Incubation Center for Future Technology of Samsung Electronics under Grant SRFC-IT1901-17 and in part by the MSIT (Ministry of Science and ICT), Korea, under the ITRC (Information Technology Research Center) support program (IITP-2020-2017-0-01637) supervised by the IITP (Institute for Information \& communications Technology Promotion).}
}

\author{*Junhan Kim, $^{\dagger}$Jian Wang, and *Byonghyo Shim, \\

$^{*}$Department of Electrical and Computer Engineering, Seoul National University, Seoul, Korea \\

$^{\dagger}$School of Data Science, Fudan University, Shanghai, China \\

Email: $^{*}$\{junhankim, bshim\}@islab.snu.ac.kr, $^{\dagger}$jian\_wang@fudan.edu.cn
}

\maketitle

\begin{abstract}
Orthogonal least squares (OLS) is a classic algorithm for sparse recovery, function approximation, and subset selection. In this paper, we analyze the performance guarantee of the OLS algorithm. Specifically, we show that OLS guarantees the exact reconstruction of any $K$-sparse vector in $K$ iterations, provided that a sensing matrix has unit $\ell_{2}$-norm columns and satisfies the restricted isometry property (RIP) of order $K+1$ with
\begin{align*}
\delta_{K+1}
&<C_{K}
= \begin{cases}
\frac{1}{\sqrt{K}}, & K=1, \\
\frac{1}{\sqrt{K+\frac{1}{4}}}, & K=2, \\
\frac{1}{\sqrt{K+\frac{1}{16}}}, & K=3, \\
\frac{1}{\sqrt{K}}, & K \ge 4.
\end{cases} 
\end{align*}
Furthermore, we show that the proposed guarantee is optimal in the sense that if $\delta_{K+1} \ge C_{K}$, then there exists a counterexample for which OLS fails the recovery.
\end{abstract}

\section{Introduction}
Orthogonal least squares (OLS) is a classic algorithm for sparse recovery, function approximation, and subset selection~\cite{chen1989orthogonal, rebollo2002optimized}. The main task of the OLS algorithm is to recover a high dimensional $K$-sparse vector $\mathbf{x} \in \mathbb{R}^{n}~(\| \mathbf{x} \|_{0} \le K \ll n)$ from a small number of measurements 
\begin{align}
\mathbf{y}
&= \mathbf{A} \mathbf{x}, \label{eq:system}
\end{align}
where $\mathbf{A} \in \mathbb{R}^{m \times n}$ $(m \ll n)$ is the sensing matrix. In order to perform this task, OLS investigates the support (the index set of nonzero elements) of $\mathbf{x}$ sequentially in a way to minimize the residual power. Specifically, in each iteration, OLS identifies the column of $\mathbf{A}$ that leads to the most significant reduction in the $\ell_{2}$-norm of a residual and then adds the index of the column to the estimated support. The vestige of indices in the enlarged support is then removed from $\mathbf{y}$, generating an updated residual for the upcoming iteration. As in~\cite{wang2017recovery, wen2017nearly, kim2018multiple, kim2019near, kim2019nearly}, we restrict our attention to the OLS algorithm running $K$ iterations in this paper. In Table~\ref{tab:OLS}, we summarize the OLS algorithm. 

\setlength{\arrayrulewidth}{1.6pt}
\begin{table}
  \centering
\caption{The OLS Algorithm} \label{tab:OLS}
\vspace{-2mm}
\begin{tabular}{@{}ll}
\hline \vspace{-5pt} \\
\textbf{~~~Input}       & sensing matrix $\mathbf{A}$, \\
						& measurement vector $\mathbf{y}$, \\
						& and sparsity $K$. \\
\textbf{~~~Initialize}  & iteration counter $k = 0$, \\
                     & estimated support ${S}^{0} = \emptyset$, \\
                     & and residual vector $\mathbf{r}^{0} = \mathbf{y}$.
                     \\
\textbf{~~~While}       & $k < K$ \ \textbf{do}\\
                     & $k = k + 1$; \\
                     & (Identify{\color{black}$^a$}) \hspace{1.2mm}$s^{k} = \underset{i \in \{1, \ldots, n\}}{\arg \min}  \| \mathbf{P}^{\bot}_{S^{k - 1} \cup \{i\}} \mathbf{y} \|_{2}^{2}$; \\
                     & (Augment) \hspace{.6mm}$S^{k} = S^{k - 1} \cup \{ s^{k} \}$; \\
                     & (Estimate) \hspace{.9mm}$\mathbf{x}^{k} = \underset{\mathbf{u}:\supp(\mathbf{u}) = S^{k}}{\arg \min} \|\mathbf{y} - \mathbf{A} \mathbf{u}\|_2$; \\
                     & (Update) \hspace{2.1mm} $\mathbf{r}^{k} = \mathbf{y} - \mathbf{A} \mathbf{x}^{k}$; \\
\textbf{~~~End}         \\
\textbf{~~~Output}  &$S^{K}$ and $\mathbf{x}^{K}$. \\
\vspace{-5pt} \\
\hline
\end{tabular}
    \begin{tablenotes}
        \item[a] \hspace{-2mm}{\color{black}$^a$}If the minimum occurs for multiple indices, {\color{black}{break the tie deterministically in favor of the first one.}}
    \end{tablenotes} \vspace{-3mm}
\end{table}
\setlength{\arrayrulewidth}{1.6pt}

As a framework to analyze the performance of the OLS algorithm, the restricted isometry property (RIP) has been widely used~\cite{wang2017recovery, wen2017nearly, kim2018multiple, kim2019near, kim2019nearly, kim2020joint}. A matrix $\mathbf{A}$ is said to obey the RIP of order $K$ if there exists a constant $\delta \in [0, 1)$ such that
\begin{align}
(1-\delta) \| \mathbf{x} \|_{2}^{2}
\le \| \mathbf{A} \mathbf{x} \|_{2}^{2}
\le (1+\delta) \| \mathbf{x} \|_{2}^{2} \label{def:RIP}
\end{align}
for any $K$-sparse vector $\mathbf{x}$~\cite{candes2005decoding}. In particular, the minimum of $\delta$ satisfying~\eqref{def:RIP} is called the RIP constant and denoted by $\delta_{K}$. In~\cite{wang2017recovery}, it has been shown that OLS guarantees the perfect recovery of any $K$-sparse vector in $K$ iterations if a sensing matrix has unit $\ell_{2}$-norm columns and obeys the RIP with $\delta_{K+1} < \frac{1}{\sqrt{K}+2}$. This condition has recently been improved to~\cite{wen2017nearly}
\begin{align}
\delta_{K+1}
&< \frac{1}{\sqrt{K+1}}. \label{eq:sufficient condition of OLS_l2-normalized sensing matrix_Wen and Wang's result}
\end{align}
On the other hand, it has been reported in~\cite{wen2017nearly} that when $K=2$, there exists a counterexample for which OLS fails the reconstruction under $\delta_{K+1} = \frac{1}{\sqrt{K+\frac{1}{4}}}$. Based on the counterexample, it has been conjectured that for all of $K$, a recovery guarantee of OLS cannot be weaker than
\begin{align} \label{eq:conjectured performance limit of OLS_Wen et al}
\delta_{K+1}
&< \frac{1}{\sqrt{K+\frac{1}{4}}}.
\end{align}

The main purpose of this paper is to bridge the gap between~\eqref{eq:sufficient condition of OLS_l2-normalized sensing matrix_Wen and Wang's result} and~\eqref{eq:conjectured performance limit of OLS_Wen et al}. To this end, we first present an improved recovery guarantee of OLS. Specifically, we show that OLS exactly recovers any $K$-sparse vector in $K$ iterations if a sensing matrix has unit $\ell_{2}$-norm columns and obeys the RIP with
\begin{align} 
\delta_{K+1}
< C_{K}
= \begin{cases}
\frac{1}{\sqrt{K}}, & K=1, \\
\frac{1}{\sqrt{K+\frac{1}{4}}}, & K=2, \\
\frac{1}{\sqrt{K+\frac{1}{16}}}, & K=3, \\
\frac{1}{\sqrt{K}}, & K \ge 4.
\end{cases} \label{eq:performance guarantee of OLS_proposed} 
\end{align}

The significance of our result lies in the fact that it not only outperforms the existing result in~\eqref{eq:sufficient condition of OLS_l2-normalized sensing matrix_Wen and Wang's result}, but also states an optimal recovery guarantee of OLS. By the optimality, we mean that if the condition is violated, then there exists a counterexample for which OLS fails the recovery. In fact, for any positive integer $K$ and constant $\delta^{*} \in [C_{K}, 1)$, there always exist an $\ell_{2}$-normalized matrix $\mathbf{A}$ with $\delta_{K+1} = \delta^{*}$ and a $K$-sparse vector $\mathbf{x}$ such that $\mathbf{x}$ cannot be recovered from $\mathbf{y}=\mathbf{A} \mathbf{x}$ by OLS.
\section{Preliminaries}

We first summarize the notations used in this paper. 
\begin{itemize}
\item $\Omega = \{ 1, \ldots, n \}$;
\item $S = \supp(\mathbf{x}) = \{ i \in \Omega : x_{i} \neq 0 \}$ is the support of $\mathbf{x}$;
\item For any $J \subseteq \Omega$, $|J|$ is the cardinality of $J$ and $S \setminus J = \{ i : i \in S, i \notin J \}$;
\item $\mathbf{x}_{J} \in \mathbb{R}^{|J|}$ is the restriction of $\mathbf{x} \in \mathbb{R}^{n}$ to the elements indexed by $J$;
\item $\mathbf{a}_{i} \in \mathbb{R}^{m}$ is the $i$-th column of $\mathbf{A}\in \mathbb{R}^{m \times n}$;
\item $\mathbf{A}_{J} \in \mathbb{R}^{m \times |J|}$ is the submatrix of $\mathbf{A}$ with the columns indexed by $J$;
\item $\text{span}(\mathbf{A}_{J})$ is the column space of $\mathbf{A}_{J}$;
\item If $\mathbf{A}_{J}$ has full column rank, then $\mathbf{A}_{J}^{\dagger} = (\mathbf{A}_{J}^{\prime} \mathbf{A}_{J})^{-1} \mathbf{A}_{J}^{\prime}$ is the pseudoinverse of $\mathbf{A}_{J}$ where $\mathbf{A}_{J}^{\prime}$ is the transpose of $\mathbf{A}_{J}$;
\item $\mathbf{0}_{d_{1} \times d_{2}}$ and $\mathbf{1}_{d_{1} \times d_{2}}$ are ($d_{1} \times d_{2}$)-dimensional matrices with entries being zeros and ones, respectively;
\item $\mathbf{Id}_{d}$ is the $d$-dimensional identity matrix;
\item $\mathbf{P}_{J} = \mathbf{A}_{J} \mathbf{A}_{J}^{\dagger}$ and $\mathbf{P}_{J}^{\perp} = \mathbf{Id}_{m} - \mathbf{P}_{J}$ are the orthogonal projections onto $\text{span}(\mathbf{A}_{J})$ and its orthogonal complement, respectively.
\end{itemize}

We next give some lemmas useful in our analysis. The first lemma is about the monotonicity of the RIP constant.

\begin{lemma}[\hspace{-0.2mm}{\cite{candes2005decoding},~\cite[Lemma 1]{dai2009subspace},~\cite{wang2012generalized}}] \label{lemma:monotonicity of RIC}
If $\mathbf{A}$ obeys the RIP of orders $K_{1}$ and $K_{2}$ $(K_{1} \le K_{2})$, then $\delta_{K_{1}} \le \delta_{K_{2}}$.
\end{lemma}

The second lemma is often called the modified RIP of a projected matrix.

\begin{lemma}[\hspace{-0.2mm}{\cite[Lemma 1]{li2015sufficient}}] \label{lemma:RIC of projected matrix}
Let $S, J \subset \Omega$. If $\mathbf{A} \in \mathbb{R}^{m \times n}$ satisfies the RIP of order $|S \cup J|$, then for any $\mathbf{x} \in \mathbb{R}^{n}$ with $\supp(\mathbf{x})=S$,
\begin{align*}
(1 - \delta_{|S \cup J|}) \| \mathbf{x}_{S \setminus J} \|_{2}^{2}
\le \| \mathbf{P}_{J}^{\perp} \mathbf{A} \mathbf{x} \|_{2}^{2}
\le (1 + \delta_{|S \cup J|}) \| \mathbf{x}_{S \setminus J} \|_{2}^{2}. 
\end{align*}
\end{lemma}

The third lemma gives an equivalent form to the identification rule of OLS in Table~\ref{tab:OLS}.

\begin{lemma}[\hspace{-0.2mm}{\cite[Theorem 1]{rebollo2002optimized}}] \label{lemma:refined identification rule of OLS}
Consider the system model in~\eqref{eq:system}. Let $S^{k}$ and $\mathbf{r}^{k}$ be the estimated support and the residual generated in the $k$-th iteration of the OLS algorithm, respectively. Then the index $s^{k+1}$ chosen in the $(k+1)$-th iteration of OLS satisfies
\begin{align}
s^{k+1}
&= \underset{j \in \Omega \setminus S^{k}}{\arg \max} \frac{| \langle \mathbf{r}^{k}, \mathbf{a}_{j} \rangle |}{\| \mathbf{P}_{S^{k}}^{\perp} \mathbf{a}_{j} \|_{2}}. \label{eq:refined identification rule of OLS}
\end{align}
\end{lemma}

One can deduce from~\eqref{eq:refined identification rule of OLS} that OLS picks a support index in the ($k+1$)-th iteration (i.e., $s^{k+1} \in S$) if and only if 
\begin{align}
\max_{j \in S \setminus S^{k}} \frac{| \langle \mathbf{r}^{k}, \mathbf{a}_{j} \rangle |}{\| \mathbf{P}_{S^{k}}^{\perp} \mathbf{a}_{j} \|_{2}}
&> \max_{j \in \Omega \setminus (S \cup S^{k})} \frac{| \langle \mathbf{r}^{k}, \mathbf{a}_{j} \rangle |}{\| \mathbf{P}_{S^{k}}^{\perp} \mathbf{a}_{j} \|_{2}}. \label{eq:correct selection for OLS}
\end{align}
To examine if OLS is successful in the $(k+1)$-th iteration, therefore, it suffices to check whether~\eqref{eq:correct selection for OLS} holds. 

The next lemma plays a key role in bounding the right-hand side of~\eqref{eq:correct selection for OLS}.

\begin{lemma} \label{lemma:primary novelty}
Consider the system model in~\eqref{eq:system} where $\mathbf{A}$ has unit $\ell_{2}$-norm columns. Let $S^{k}$ be a subset of $S = \supp(\mathbf{x})$ and $\mathbf{r}^{k} = \mathbf{P}_{S^{k}}^{\perp} \mathbf{A} \mathbf{x}$. If $\mathbf{A}$ obeys the RIP of order $K+1$ with
\begin{align} \label{eq:primary novelty_RIP condition}
\delta_{K+1}
&\le \frac{1}{2},
\end{align}
then
\begin{align}
\max_{j \in \Omega \setminus S} \frac{| \langle \mathbf{r}^{k}, \mathbf{a}_{j} \rangle |}{\| \mathbf{P}_{S^{k}}^{\perp} \mathbf{a}_{j} \|_{2}}
&\le \frac{\delta_{K+1}\| \mathbf{r}^{k} \|_{2}^{2}}{\| \mathbf{x}_{S \setminus S^{k}} \|_{2}}. \label{eq:primary novelty}
\end{align}
\end{lemma}

\proof
Let $j \in \Omega \setminus S$ and $t = \| \mathbf{x}_{S \setminus S^{k}} \|_{2} | \langle \mathbf{r}^{k}, \mathbf{a}_{j} \rangle | / \| \mathbf{P}_{S^{k}}^{\perp} \mathbf{a}_{j} \|_{2}$.
Then, what we need to show is
\begin{align} \label{eq:proof_primary novelty_target}
t 
&\le \delta_{K+1} \| \mathbf{r}^{k} \|_{2}^{2}
\end{align} 
under~\eqref{eq:primary novelty_RIP condition}. Let $\mathbf{\Psi} = \mathbf{P}_{S^{k}}^{\perp} [ \mathbf{A}_{S \setminus S^{k}} \ \mathbf{a}_{j} ]$ and
\begin{align*} 
\mathbf{u}
= \begin{bmatrix}
\mathbf{x}_{S \setminus S^{k}} \\
-\frac{\sgn(\mathbf{a}_{j}^{\prime} \mathbf{r}^{k}) \| \mathbf{x}_{S \setminus S^{k}} \|_{2}}{\| \mathbf{P}_{S^{k}}^{\perp} \mathbf{a}_{j} \|_{2}}
\end{bmatrix},~
\mathbf{v}
= \begin{bmatrix}
\mathbf{x}_{S \setminus S^{k}} \\
\frac{\sgn(\mathbf{a}_{j}^{\prime} \mathbf{r}^{k}) \| \mathbf{x}_{S \setminus S^{k}} \|_{2}}{\| \mathbf{P}_{S^{k}}^{\perp} \mathbf{a}_{j} \|_{2}}
\end{bmatrix}, 
\end{align*}
where $\sgn(\cdot)$ is the signum function. By noting that $\mathbf{r}^{k} = \mathbf{P}_{S^{k}}^{\perp} \mathbf{A}_{S \setminus S^{k}} \mathbf{x}_{S \setminus S^{k}}$ and $\mathbf{P}_{S^{k}}^{\perp} = (\mathbf{P}_{S^{k}}^{\perp})^{\prime} = (\mathbf{P}_{S^{k}}^{\perp})^{2}$, we have
\begin{subequations}
\begin{align}
\| \mathbf{\Psi} \mathbf{u} \|_{2}^{2}
&= \| \mathbf{r}^{k} \|_{2}^{2} - 2t + \| \mathbf{x}_{S \setminus S^{k}} \|_{2}^{2}, \label{eq:pf of prop 1_1_(a)_K>3} \\
\| \mathbf{\Psi} \mathbf{v} \|_{2}^{2}
&=\| \mathbf{r}^{k} \|_{2}^{2} + 2t + \| \mathbf{x}_{S \setminus S^{k}} \|_{2}^{2}. \label{eq:pf of prop 1_2_(a)_K>3}
\end{align}
\end{subequations}
Also, from Lemma~\ref{lemma:RIC of projected matrix}, we have
\begin{subequations}
\begin{align}
\| \mathbf{\Psi} \mathbf{u} \|_{2}^{2}
&\ge (1 - \delta_{K+1}) \left ( 1 + \frac{1}{\| \mathbf{P}_{S^{k}}^{\perp} \mathbf{a}_{j} \|_{2}^{2}} \right )\| \mathbf{x}_{S \setminus S^{k}} \|_{2}^{2}, \label{eq:pf of prop 1_1_(b)_K>3} \\
\| \mathbf{\Psi} \mathbf{v} \|_{2}^{2}
&\le (1 + \delta_{K+1}) \left ( 1 + \frac{1}{\| \mathbf{P}_{S^{k}}^{\perp} \mathbf{a}_{j} \|_{2}^{2}} \right ) \| \mathbf{x}_{S \setminus S^{k}} \|_{2}^{2}. \label{eq:pf of prop 1_2_(b)_K>3}
\end{align}
\end{subequations}
By combining~\eqref{eq:pf of prop 1_1_(a)_K>3} and~\eqref{eq:pf of prop 1_1_(b)_K>3}, we obtain
\begin{align} 
\| \mathbf{r}^{k} \|_{2}^{2} - 2t
&\ge \left ( 1 - \left ( 1 + \| \mathbf{P}_{S^{k}}^{\perp} \mathbf{a}_{j} \|_{2}^{2} \right ) \delta_{K+1} \right ) \frac{\| \mathbf{x}_{S \setminus S^{k}} \|_{2}^{2}}{\| \mathbf{P}_{S^{k}}^{\perp} \mathbf{a}_{j} \|_{2}^{2}}. \label{eq:main_1}
\end{align}
Also, by combining~\eqref{eq:pf of prop 1_2_(a)_K>3} and~\eqref{eq:pf of prop 1_2_(b)_K>3}, we obtain
\begin{align} 
\| \mathbf{r}^{k} \|_{2}^{2} + 2t
&\le \left ( 1 + \left ( 1 + \| \mathbf{P}_{S^{k}}^{\perp} \mathbf{a}_{j} \|_{2}^{2} \right ) \delta_{K+1} \right ) \frac{\| \mathbf{x}_{S \setminus S^{k}} \|_{2}^{2}}{\| \mathbf{P}_{S^{k}}^{\perp} \mathbf{a}_{j} \|_{2}^{2}}. \label{eq:main_2}
\end{align}
Note that since $\mathbf{A}$ has unit $\ell_{2}$-norm columns, $\| \mathbf{P}_{S^{k}}^{\perp} \mathbf{a}_{j} \|_{2}^{2} \le \| \mathbf{a}_{j} \|_{2}^{2} = 1$ and thus we have
\begin{align*}
1 - \left ( 1 + \| \mathbf{P}_{S^{k}}^{\perp} \mathbf{a}_{j} \|_{2}^{2} \right ) \delta_{K+1}
\ge 1-2\delta_{K+1}
\overset{(a)}{\ge} 0,
\end{align*}
where (a) is from~\eqref{eq:primary novelty_RIP condition}. Then, by combining~\eqref{eq:main_1} and~\eqref{eq:main_2}, we have
\begin{align*}
\lefteqn{(\| \mathbf{r}^{k} \|_{2}^{2} -2t) \left ( 1 + \left ( 1 + \| \mathbf{P}_{S^{k}}^{\perp} \mathbf{a}_{j} \|_{2}^{2} \right ) \delta_{K+1} \right )} \nonumber \\
&~~~~~~~~~\ge \left ( \| \mathbf{r}^{k} \|_{2}^{2} + 2t \right ) \left ( 1 - \left ( 1 + \| \mathbf{P}_{S^{k}}^{\perp} \mathbf{a}_{j} \|_{2}^{2} \right ) \delta_{K+1} \right ),
\end{align*}
which is equivalent to
\begin{align*}
2t
&\le ( 1 + \| \mathbf{P}_{S^{k}}^{\perp} \mathbf{a}_{j} \|_{2}^{2} ) \delta_{K+1} \| \mathbf{r}^{k} \|_{2}^{2}. 
\end{align*}
Finally, by exploiting $\| \mathbf{P}_{S^{k}}^{\perp} \mathbf{a}_{j} \|_{2}^{2} \le 1$, we obtain the desired result~\eqref{eq:proof_primary novelty_target}. \endproof

\begin{remark}
The bound in~\eqref{eq:primary novelty} is tight in the sense that the equality of~\eqref{eq:primary novelty} is attainable. To see this, we consider the following example: 
\begin{align*}
\mathbf{x}
= \begin{bmatrix}
0 \\
\mathbf{1}_{K \times 1}
\end{bmatrix}~\andd~ 
\mathbf{A} = \begin{bmatrix}
\sqrt{\frac{K-1}{K}} & \mathbf{0}_{1 \times K} \\
\frac{1}{K}\mathbf{1}_{K \times 1} & \mathbf{Id}_{K}
\end{bmatrix}.
\end{align*}
We take a look at the left- and right-hand sides of~\eqref{eq:primary novelty} when $k=0$. Since $\mathbf{y} = \mathbf{A} \mathbf{x} = \mathbf{x}$, the left-hand side of~\eqref{eq:primary novelty} is 
\begin{align*}
\max_{j \in \Omega \setminus S} \frac{| \langle \mathbf{r}^{0}, \mathbf{a}_{j} \rangle |}{\| \mathbf{P}_{S^{0}}^{\perp} \mathbf{a}_{j} \|_{2}}
&= | \langle \mathbf{y}, \mathbf{a}_{1} \rangle |
= 1.
\end{align*}
Furthermore, since the RIP constant $\delta_{K+1}$ of $\mathbf{A}$ is $\delta_{K+1} = \frac{1}{\sqrt{K}}$ (see~\cite[p.3655]{mo2012remarks}), the right-hand side of~\eqref{eq:primary novelty} is also
\begin{align*}
\frac{\delta_{K+1} \| \mathbf{r}^{0} \|_{2}^{2}}{\| \mathbf{x}_{S \setminus S^{0}} \|_{2}}
&= \frac{\delta_{K+1} \| \mathbf{y} \|_{2}^{2}}{\| \mathbf{x}_{S} \|_{2}}
=1.
\end{align*} 
Therefore, the bound in~\eqref{eq:primary novelty} is tight.
\end{remark}

\begin{remark} \label{rmk:novelty of Lemma}
Lemma~\ref{lemma:primary novelty} is motivated by~\cite[Lemma 4]{wen2017nearly}, where the inequality 
\begin{align}
\max_{j \in \Omega \setminus S} | \langle  \mathbf{r}^{k}, \mathbf{a}_{j} \rangle |
&\le \frac{\|  \mathbf{r}^{k} \|_{2}^{2}}{\sqrt{\alpha}\| \mathbf{x}_{S \setminus S^{k}} \|_{2}} \label{eq:Wen's result}
\end{align}
was established under
\begin{align} \label{eq:Wen's result_RIP condition}
\delta_{K+1} 
&\le \frac{1}{\sqrt{\alpha+1}}.
\end{align}
We note that our result in Lemma~\ref{lemma:primary novelty} outperforms this result in the following aspects:
\begin{enumerate}[i)]
\item A tighter upper bound of $\max_{j \in \Omega \setminus S} | \langle \mathbf{r}^{k}, \mathbf{a}_{j} \rangle |$ can be established using Lemma~\ref{lemma:primary novelty}. By applying~\cite[Lemma 4]{wen2017nearly} with $\alpha = 3$, we have
\begin{align} \label{eq:Wen's inequality_alpha=3}
\max_{j \in \Omega \setminus S} | \langle \mathbf{r}^{k}, \mathbf{a}_{j} \rangle |
&\le \frac{\| \mathbf{r}^{k} \|_{2}^{2}}{\sqrt{3}\|\mathbf{x}_{S \setminus S^{k}} \|_{2}}
\end{align}
under $\delta_{K+1} \le \frac{1}{2}$. Under the same condition on $\delta_{K+1}$, it can be deduced from Lemma~\ref{lemma:primary novelty} that
\begin{align} 
\max_{j \in \Omega \setminus S} | \langle \mathbf{r}^{k}, \mathbf{a}_{j} \rangle |
&\overset{(a)}{\le} \max_{j \in \Omega \setminus S} \frac{| \langle \mathbf{r}^{k}, \mathbf{a}_{j} \rangle |}{\| \mathbf{P}_{S^{k}}^{\perp} \mathbf{a}_{j} \|_{2}} \nonumber \\
&\hspace{.37mm}\le \frac{\delta_{K+1}\| \mathbf{r}^{k} \|_{2}^{2}}{\| \mathbf{x}_{S \setminus S^{k}} \|_{2}} \nonumber \\
&\hspace{.37mm}\le \frac{\| \mathbf{r}^{k} \|_{2}^{2}}{2\| \mathbf{x}_{S \setminus S^{k}} \|_{2}}, \label{eq:comparison with Wen's inequality_alpha=3}
\end{align}
where (a) is because $\| \mathbf{P}_{S^{k}}^{\perp} \mathbf{a}_{j} \|_{2} \le \| \mathbf{a}_{j} \|_{2} = 1$ for each of $j \in \Omega \setminus S$. Clearly, the bound in~\eqref{eq:comparison with Wen's inequality_alpha=3} is tighter than that in~\eqref{eq:Wen's inequality_alpha=3} by the factor of $2 / \sqrt{3}$.

\item In~\cite{wen2017nearly}, by putting $\alpha = |S \setminus S^{k}| / \| \mathbf{P}_{S^{k}}^{\perp} \mathbf{a}_{j} \|_{2}^{2}$ into~\eqref{eq:Wen's result}, the inequality
\begin{align}
\max_{j \in \Omega \setminus S} \frac{| \langle \mathbf{r}^{k}, \mathbf{a}_{j} \rangle |}{\| \mathbf{P}_{S^{k}}^{\perp} \mathbf{a}_{j} \|_{2}}
&< \frac{\| \mathbf{r}^{k} \|_{2}^{2}}{\sqrt{|S \setminus S^{k}|}\| \mathbf{x}_{S \setminus S^{k}} \|_{2}} \label{eq:Wen's result_OLS version}
\end{align}
was established under
\begin{align} \label{eq:Wen's result_OLS version_RIP condition}
\delta_{K+1}
&< \frac{1}{\sqrt{K+1}}.
\end{align} 
We mention that the inequality~\eqref{eq:Wen's result_OLS version} can be established under a weaker RIP condition using Lemma~\ref{lemma:primary novelty}. Specifically, when $K \ge 4$,~\eqref{eq:Wen's result_OLS version} can be obtained from Lemma~\ref{lemma:primary novelty} under 
\begin{align} \label{eq:comparison with Wen's result_main lemma_OLS version_RIP condition}
\delta_{K+1} 
&< \frac{1}{\sqrt{K}}
\end{align}
because
\begin{align*}
\max_{j \in \Omega \setminus S} \frac{| \langle \mathbf{r}^{k}, \mathbf{a}_{j} \rangle |}{\| \mathbf{P}_{S^{k}}^{\perp} \mathbf{a}_{j} \|_{2}}
&\le \frac{\delta_{K+1} \| \mathbf{r}^{k} \|_{2}^{2}}{\| \mathbf{x}_{S \setminus S^{k}} \|_{2}} \\
&< \frac{\| \mathbf{r}^{k} \|_{2}^{2}}{\sqrt{K}\| \mathbf{x}_{S \setminus S^{k}} \|_{2}} \\
&\le \frac{\| \mathbf{r}^{k} \|_{2}^{2}}{\sqrt{|S \setminus S^{k}|}\| \mathbf{x}_{S \setminus S^{k}} \|_{2}}.
\end{align*} 
One can see that~\eqref{eq:comparison with Wen's result_main lemma_OLS version_RIP condition} is less restrictive than~\eqref{eq:Wen's result_OLS version_RIP condition}.
\end{enumerate}
\end{remark}

\section{Exact Sparse Recovery with OLS}

In this section, we present an optimal recovery guarantee for the OLS algorithm. First, we present a sufficient condition ensuring the success of OLS.

\begin{theorem} \label{thm:sufficient condition of OLS}
Let $\mathbf{A} \in \mathbb{R}^{m \times n}$ be a matrix having unit $\ell_{2}$-norm columns. If $\mathbf{A}$ obeys the RIP of order $K+1$ with~\eqref{eq:performance guarantee of OLS_proposed}, then the OLS algorithm exactly reconstructs any $K$-sparse vector $\mathbf{x} \in \mathbb{R}^{n}$ from its samples $\mathbf{y} = \mathbf{A} \mathbf{x}$ in $K$ iterations.
\end{theorem}

\proof
We show that the OLS algorithm picks a support index in each iteration. In other words, we show that $S^{k} \subset S$ for all of $k \in \{ 0, \ldots, K \}$. In doing so, we have $S^{K} = S$, and thus OLS can recover $\mathbf{x}$ accurately:
\begin{align*}
(\mathbf{x}^{K})_{S^{K}} 
= \mathbf{A}_{S^{K}}^{\dagger} \mathbf{y}
= \mathbf{A}_{S^{K}}^{\dagger} \mathbf{A}_{S} \mathbf{x}_{S}
= \mathbf{A}_{S}^{\dagger} \mathbf{A}_{S} \mathbf{x}_{S}
= \mathbf{x}_{S}. 
\end{align*} 
First, we consider the case for $k=0$. This case is trivial since $S^{0} = \emptyset \subset S$. Next, we assume that $S^{k} \subset S$ for some integer $k~(0 \le k < K)$ and then show that OLS picks a support index in the $(k+1)$-th iteration. As mentioned, $s^{k+1} \in S$ if and only if the condition~\eqref{eq:correct selection for OLS} holds. Since the left-hand side of~\eqref{eq:correct selection for OLS} satisfies~\cite[Proposition 2]{wen2017nearly}
\begin{align*}
\max_{j \in S \setminus S^{k}} \frac{| \langle \mathbf{r}^{k}, \mathbf{a}_{j} \rangle |}{\| \mathbf{P}_{S^{k}}^{\perp} \mathbf{a}_{j} \|_{2}}
&\ge \frac{\| \mathbf{r}^{k} \|_{2}^{2}}{\sqrt{K-k} \| \mathbf{x}_{S \setminus S^{k}} \|_{2}},
\end{align*}
it suffices to show that
\begin{align*} 
\max_{j \in \Omega \setminus S} \frac{| \langle \mathbf{r}^{k}, \mathbf{a}_{j} \rangle |}{\| \mathbf{P}_{S^{k}}^{\perp} \mathbf{a}_{j} \|_{2}}
&< \frac{\| \mathbf{r}^{k} \|_{2}^{2}}{\sqrt{K-k} \| \mathbf{x}_{S \setminus S^{k}} \|_{2}}.
\end{align*}
Towards this end, we consider three cases: (i) $K=1$, (ii) $K \in \{ 2, 3 \}$, and (iii) $K \ge 4$.

\begin{itemize}
\item[(i)] $K=1$

In this case, $k=0$ and thus we need to show that 
\begin{align} \label{eq:target inequality_K=1}
\max_{j \in \Omega \setminus S} | \langle \mathbf{y}, \mathbf{a}_{j} \rangle |
&< \frac{\| \mathbf{y} \|_{2}^{2}}{\| \mathbf{x}_{S} \|_{2}}.
\end{align}
Without loss of generality, we assume that $\mathbf{x} = [c \ \mathbf{0}_{(n-1) \times 1}]^{\prime}$ for some $c \in \mathbb{R} \setminus \{ 0 \}$. Then, $\mathbf{y} = c \mathbf{a}_{1}$ and thus the right-hand side of~\eqref{eq:target inequality_K=1} is simply $|c|$. As a result, it suffices to show that
\begin{align} \label{eq:target inequality_K=1_modification}
\max_{j \in \Omega \setminus S} | \langle \mathbf{y}, \mathbf{a}_{j} \rangle |
&< |c|.
\end{align} 
Let $j \in \Omega \setminus S$ and $\theta$ be the angle between $\mathbf{a}_{1}$ and $\mathbf{a}_{j}$ ($0 \le \theta \le \pi$). Then, by~\cite[Lemma 2.1]{chang2014improved}, we have $| \cos \theta | \le \delta_{2}$ and hence
\begin{align*} 
| \langle \mathbf{y}, \mathbf{a}_{j} \rangle |
= | \langle c \mathbf{a}_{1}, \mathbf{a}_{j} \rangle |
= |c \cos \theta|
\le |c| \delta_{2}.
\end{align*}
Using this together with~\eqref{eq:performance guarantee of OLS_proposed}, we obtain the desired result in~\eqref{eq:target inequality_K=1_modification}.

\item[(ii)] $K \in \{ 2, 3 \}$

Let $j \in \Omega \setminus S$ and
$$q= \frac{\sqrt{K-k} \| \mathbf{x}_{S \setminus S^{k}} \|_{2} | \langle \mathbf{r}^{k}, \mathbf{a}_{j} \rangle |}{\| \mathbf{P}_{S^{k}}^{\perp} \mathbf{a}_{j} \|_{2}}.$$
Then, our task is to show that 
\begin{align} \label{eq:proof of Theorem 1_target_K=2,3}
q 
&< \| \mathbf{r}^{k} \|_{2}^{2}.
\end{align}
Let $\mathbf{\Psi}
= \mathbf{P}_{S^{k}}^{\perp} [\mathbf{A}_{S \setminus S^{k}} \ \mathbf{a}_{j}]$ and 
\begin{align*} 
\mathbf{w}
&= \begin{bmatrix}
\mathbf{x}_{S \setminus S^{k}} \\
-\frac{\text{sgn}(\mathbf{a}_{j}^{\prime} \mathbf{r}^{k}) \sqrt{K-k} \| \mathbf{x}_{S \setminus S^{k}} \|_{2}}{2\| \mathbf{P}_{S^{k}}^{\perp} \mathbf{a}_{j} \|_{2}}
\end{bmatrix}.
\end{align*}
Then, we have
\begin{align}
\| \mathbf{\Psi} \mathbf{w} \|_{2}^{2}
&= \| \mathbf{r}^{k} \|_{2}^{2} - q  + \frac{(K-k)\| \mathbf{x}_{S \setminus S^{k}} \|_{2}^{2}}{4}. \label{eq:pf of prop 1_1_(a)_K=2,3}
\end{align}
Also, from Lemma~\ref{lemma:RIC of projected matrix}, we have
\begin{align}
\| \mathbf{\Psi} \mathbf{w} \|_{2}^{2}
&\hspace{.37mm}\ge (1 - \delta_{K+1}) \left ( 1 + \frac{K-k}{4\| \mathbf{P}_{S^{k}}^{\perp} \mathbf{a}_{j} \|_{2}^{2}} \right ) \| \mathbf{x}_{S \setminus S^{k}} \|_{2}^{2} \nonumber \\
&\overset{(a)}{\ge} (1 - \delta_{K+1}) \left ( 1 + \frac{K-k}{4} \right ) \| \mathbf{x}_{S \setminus S^{k}} \|_{2}^{2}, \label{eq:pf of prop 1_1_(b)_K=2,3}
\end{align}
where (a) is because $\| \mathbf{P}_{S^{k}}^{\perp} \mathbf{a}_{j} \|_{2} \le \| \mathbf{a}_{j} \|_{2} = 1$. By combining~\eqref{eq:pf of prop 1_1_(a)_K=2,3} and~\eqref{eq:pf of prop 1_1_(b)_K=2,3}, we obtain
\begin{align*}
\| \mathbf{r}^{k} \|_{2}^{2} - q
&\hspace{.37mm} \ge \left ( 1 - \left ( 1 + \frac{K-k}{4} \right ) \delta_{K+1} \right ) \| \mathbf{x}_{S \setminus S^{k}} \|_{2}^{2} \\
&\overset{(a)}{>} \left ( 1 - \left ( 1 + \frac{K}{4} \right ) C_{K} \right ) \| \mathbf{x}_{S \setminus S^{k}} \|_{2}^{2} \\
&\overset{(b)}{=} 0,
\end{align*}
where (a) follows from~\eqref{eq:performance guarantee of OLS_proposed} and (b) is because $$\left ( 1 + \frac{K}{4} \right ) C_{K} = 1$$
for each of $K \in \{ 2, 3 \}$ (see~\eqref{eq:performance guarantee of OLS_proposed}). Thus, we have~\eqref{eq:proof of Theorem 1_target_K=2,3}, which is the desired result.

\item[(iii)] $K \ge 4$

In this case, the RIP constant $\delta_{K+1}$ of $\mathbf{A}$ satisfies
\begin{align*}
\delta_{K+1} \overset{(a)}{<} \frac{1}{\sqrt{K}} \le \frac{1}{2},
\end{align*}
where (a) is from~\eqref{eq:performance guarantee of OLS_proposed}. Then, by applying Lemma~\ref{lemma:primary novelty}, we obtain
\begin{align*}
\max_{j \in \Omega \setminus S} \frac{| \langle \mathbf{r}^{k}, \mathbf{a}_{j} \rangle |}{\| \mathbf{P}_{S^{k}}^{\perp} \mathbf{a}_{j} \|_{2}}
&\hspace{.37mm}\le \frac{\delta_{K+1}\| \mathbf{r}^{k} \|_{2}^{2}}{\| \mathbf{x}_{S \setminus S^{k}} \|_{2}} \\
&\overset{(a)}{<} \frac{\| \mathbf{r}^{k} \|_{2}^{2}}{\sqrt{K} \| \mathbf{x}_{S \setminus S^{k}} \|_{2}} \\
&\hspace{0.37mm}\le \frac{\| \mathbf{r}^{k} \|_{2}^{2}}{\sqrt{K-k}\| \mathbf{x}_{S \setminus S^{k}} \|_{2}}, 
\end{align*}
where (a) is from~\eqref{eq:performance guarantee of OLS_proposed}. This completes the proof.
\endproof
\end{itemize}

There have been previous studies analyzing the recovery guarantee of the OLS algorithm~\cite{wang2017recovery, wen2017nearly, kim2018multiple, kim2019near}. So far, the best result states that~\eqref{eq:sufficient condition of OLS_l2-normalized sensing matrix_Wen and Wang's result} is sufficient to guarantee the exact reconstruction~\cite[Theorem 1]{wen2017nearly}. Clearly, the proposed performance guarantee~\eqref{eq:performance guarantee of OLS_proposed} is less restrictive than~\eqref{eq:sufficient condition of OLS_l2-normalized sensing matrix_Wen and Wang's result} in all sparsity region.

We now demonstrate that the proposed guarantee~\eqref{eq:performance guarantee of OLS_proposed} is optimal. This argument is established by showing that if $\delta_{K+1} \ge C_{K}$, then there exists a counterexample for which OLS fails the recovery.

\begin{theorem}\label{thm:necessary condition of OLS}
For any positive integer $K$ and constant 
\begin{align} \label{eq:recovery limit of OLS}
\delta^{*} 
&\in [C_{K}, 1),
\end{align} 
there always exist an $\ell_{2}$-normalized matrix $\mathbf{A}$ with $\delta_{K+1} = \delta^{*}$ and a $K$-sparse vector $\mathbf{x}$ such that the OLS algorithm fails to recover $\mathbf{x}$ from $\mathbf{y} = \mathbf{A} \mathbf{x}$ in $K$ iterations.
\end{theorem}

\proof
It is enough to consider the case where $K>1$ since there is no $\delta^{*}$ satisfying $\delta^{*} \in [C_{K}, 1)$ when $K=1$. In our proof, we consider three cases: (i) $K=2$, (ii) $K=3$, and (iii) $K \ge 4$.

\begin{itemize}
\item[(i)] $K=2$

We consider $\mathbf{x} = [0 \ 1 \ 1]^{\prime}$ and an $\ell_{2}$-normalized matrix $\mathbf{A}$ satisfying
\begin{align*} 
\mathbf{A}^{\prime} \mathbf{A} = \begin{bmatrix}
1 & \frac{\delta^{*}}{2} & \frac{\delta^{*}}{2} \\
\frac{\delta^{*}}{2} & 1 & -\frac{\delta^{*}}{2} \\
\frac{\delta^{*}}{2} & -\frac{\delta^{*}}{2} & 1
\end{bmatrix}.
\end{align*}
First, we compute the RIP constant of $\mathbf{A}$. Note that the eigenvalues $\lambda_{1},  \lambda_{2}, \lambda_{3}$ of $\mathbf{A}^{\prime} \mathbf{A}$ are 
\begin{align*} 
\lambda_{1} = \lambda_{2} = 1 + \frac{\delta^{*}}{2},~
\lambda_{3} = 1 - \delta^{*}.
\end{align*}
Then, by exploiting the connection between the eigenvalues of $\mathbf{A}^{\prime} \mathbf{A}$ and the RIP constant of $\mathbf{A}$~\cite[Remark 1]{dai2009subspace}, we have
\begin{align*}
\delta_{3} = \max_{i \in \{ 1, 2, 3 \}} | \lambda_{i} - 1| = \delta^{*}.
\end{align*}
In short, $\mathbf{A}$ is an $\ell_{2}$-normalized matrix satisfying the RIP with $\delta_{K+1} = \delta^{*}$. We now take a look at the first iteration of the OLS algorithm. Note that 
\begin{align*}
| \langle \mathbf{y}, \mathbf{a}_{j} \rangle |
&= | (\mathbf{A}^{\prime} \mathbf{A} \mathbf{x})_{j}|
= \begin{cases}
\delta^{*}, & j=1, \\
1 - \frac{\delta^{*}}{2}, & j=2, 3,
\end{cases}
\end{align*}
and 
$$\delta^{*} \ge C_{2} = \left . \frac{1}{\sqrt{K+\frac{1}{4}}} \right |_{K=2} = \frac{2}{3}$$
by~\eqref{eq:recovery limit of OLS}. Thus, the index $s^{1}$ chosen in the first iteration would be\footnote{When $\delta^{*} = \frac{2}{3}$, $s^{1}=1$ by the tie-breaking rule of OLS (see Table~\ref{tab:OLS}).}
$$s^{1} 
= \underset{j \in \{ 1, 2, 3 \}}{\arg \max} \hspace{.5mm} \frac{| \langle \mathbf{r}^{0}, \mathbf{a}_{j} \rangle |}{\| \mathbf{P}_{S^{0}}^{\perp} \mathbf{a}_{j} \|_{2}} 
= \underset{j \in \{ 1, 2, 3 \}}{\arg \max} \hspace{.5mm} | \langle \mathbf{y}, \mathbf{a}_{j} \rangle |
= 1,$$
and hence OLS cannot recover $\mathbf{x}$ in $K(=2)$ iterations.

\item[(ii)] $K=3$

In this case, let $\mathbf{x} = [0 \ 1 \ 1 \ 1]^{\prime}$ and 
\begin{align*}
\mathbf{A}^{\prime} \mathbf{A}
= \begin{bmatrix}
1 & \frac{\delta^{*}}{2} & \frac{\delta^{*}}{2} & \frac{\delta^{*}}{2} \\
\frac{\delta^{*}}{2} & 1 & -\frac{\delta^{*}}{8} & -\frac{\delta^{*}}{8} \\ 
\frac{\delta^{*}}{2} & -\frac{\delta^{*}}{8} & 1 & -\frac{\delta^{*}}{8} \\
\frac{\delta^{*}}{2} & -\frac{\delta^{*}}{8} & -\frac{\delta^{*}}{8} & 1
\end{bmatrix}.
\end{align*}
By taking similar steps as in case (i), one can show that $\delta_{4} = \delta^{*}$. Also, note that
\begin{align*}
| \langle \mathbf{y}, \mathbf{a}_{j} \rangle |
&= | (\mathbf{A}^{\prime} \mathbf{A} \mathbf{x})_{j}| 
= \begin{cases}
\frac{3\delta^{*}}{2}, & j=1, \\
1 - \frac{\delta^{*}}{4}, & j= 2, 3, 4,
\end{cases}
\end{align*}
and $\delta^{*} \ge \frac{4}{7}$ by~\eqref{eq:recovery limit of OLS}. Then, by~\eqref{eq:refined identification rule of OLS}, OLS would pick the first index in the first iteration (i.e., $s^{1}=1$) and thus cannot recover $\mathbf{x}$ in $K(=3)$ iterations.

\item[(iii)] $K \ge 4$

In this case, let 
\begin{align*} 
\mathbf{x}
= \begin{bmatrix}
0 \\
\mathbf{1}_{K \times 1}
\end{bmatrix},~
\mathbf{A}^{\prime} \mathbf{A}
=\begin{bmatrix}
1 & \frac{\delta^{*}}{\sqrt{K}} \mathbf{1}_{1 \times K} \\
\frac{\delta^{*}}{\sqrt{K}} \mathbf{1}_{K \times 1} & \mathbf{Id}_{K} 
\end{bmatrix}.
\end{align*}
By taking similar steps as in case (i), one can easily show that $\delta_{K+1} = \delta^{*}$. Also, by noting that 
\begin{align*}
| \langle \mathbf{y}, \mathbf{a}_{j} \rangle |
&= | (\mathbf{A}^{\prime} \mathbf{A} \mathbf{x})_{j} |
=\begin{cases}
\sqrt{K} \delta^{*}, & j = 1, \\
1, & j = 2, \ldots, K
\end{cases}
\end{align*}
and $\delta^{*} \ge \frac{1}{\sqrt{K}}$, we know that the first index would be chosen in the first iteration of OLS. Therefore, OLS cannot recover $\mathbf{x}$ in $K$ iterations. \endproof
\end{itemize}

\begin{remark}
A recovery limit of OLS over which the exact recovery is not uniformly ensured has also been studied in~\cite{wen2017nearly}. The main difference of our work is that our result holds for all $K$ and each of $\delta_{K+1} \in [C_{K}, 1)$. This is in contrast to the result in~\cite{wen2017nearly}, which is limited to the case where $K=2$ and $\delta_{K+1} = C_{K}$. 
\end{remark}

Theorem~\ref{thm:necessary condition of OLS} indicates that a performance guarantee of OLS cannot be less restrictive than 
$$\delta_{K+1} < C_{K}.$$
Combining this with the result in Theorem~\ref{thm:sufficient condition of OLS}, we conclude that the proposed guarantee~\eqref{eq:performance guarantee of OLS_proposed} is optimal.

\section{Conclusion}

In this paper, we analyzed the performance guarantee of the OLS algorithm. First, we showed that if the columns of a sensing matrix are $\ell_{2}$-normalized, then OLS ensures the exact recovery of any $K$-sparse vector in $K$ iterations under~\eqref{eq:performance guarantee of OLS_proposed}. 
\begin{align*} 
\delta_{K+1}
< C_{K}
= \begin{cases}
\frac{1}{\sqrt{K}}, & K=1, \\
\frac{1}{\sqrt{K+\frac{1}{4}}}, & K=2, \\
\frac{1}{\sqrt{K+\frac{1}{16}}}, & K=3, \\
\frac{1}{\sqrt{K}}, & K \ge 4.
\end{cases} 
\end{align*}
We next showed the optimality of the proposed guarantee by providing a counterexample for which OLS fails recovery under $\delta_{K+1} \ge C_{K}$.




\begin{thebibliography}{10}

\bibitem{chen1989orthogonal}
S.~Chen, S.~A.~Billings, and W.~Luo,
\newblock ``Orthogonal least squares methods and their application to non-linear system identification,''
\newblock {\em Int. J. Control}, vol. 50, no. 5, pp. 1873--1896, 1989.

\bibitem{rebollo2002optimized}
L.~Rebollo-Neira, D.~Lowe,
\newblock ``Optimized orthogonal matching pursuit approach,''
\newblock {\em IEEE Signal Process. Lett.}, vol. 9, no. 4, pp. 137--140, Apr. 2002.

\bibitem{wang2017recovery}
J.~Wang and P.~Li,
\newblock ``Recovery of sparse signals using multiple orthogonal least squares,''
\newblock {\em IEEE Trans. Signal Process.}, vol. 65, no. 8, pp. 2049--2062, Apr. 2017.

\bibitem{wen2017nearly}
J.~Wen, J.~Wang, and Q.~Zhang,
\newblock ``Nearly optimal bounds for orthogonal least squares,''
\newblock {\em IEEE Trans. Signal Process.}, vol. 65, no. 20, pp. 5347--5356, Oct. 2017.

\bibitem{kim2018multiple}
J.~Kim and B.~Shim,
\newblock ``Multiple orthogonal least squares for joint sparse recovery,''
\newblock in {\em Proc. IEEE Int. Symp. Inf. Theory}, Vail, CO, USA, Jun. 2018, pp. 61--65.

\bibitem{kim2019near}
J.~Kim and B.~Shim,
\newblock ``A near-optimal restricted isometry condition of multiple orthogonal least squares,''
\newblock {\em IEEE Access}, vol. 7, no. 1, pp. 46822--46830, Mar. 2019.

\bibitem{kim2019nearly}
J.~Kim, J.~Wang and B.~Shim,
\newblock ``Nearly optimal restricted isometry condition for rank aware order recursive matching pursuit,''
\newblock {\em IEEE Trans. Signal Process.}, vol. 67, no. 17, pp. 4449--4463, Sep. 2019.

\bibitem{kim2020joint}
J.~Kim, J.~Wang, L.~T.~Nguyen, and B.~Shim,
\newblock ``Joint sparse recovery using signal space matching pursuit,''
\newblock {\em IEEE Trans. Inf. Theory}, Early Access, 2020. (DOI:10.1109/TIT.2020.2986917)

\bibitem{candes2005decoding}
E.~J. Cand{\`e}s and T.~Tao,
\newblock ``{Decoding by linear programming},''
\newblock {\em IEEE Trans. Inf. Theory}, vol. 51, no. 12, pp. 4203--4215, Dec. 2005.

\bibitem{dai2009subspace}
W.~Dai and O.~Milenkovic,
\newblock ``{Subspace pursuit for compressive sensing signal reconstruction},''
\newblock {\em IEEE Trans. Inf. Theory}, vol. 55, no. 5, pp. 2230--2249, May 2009.
  
\bibitem{wang2012generalized}
J.~Wang, S.~Kwon, and B.~Shim,
\newblock ``Generalized orthogonal matching pursuit,''
\newblock {\em IEEE Trans. Signal Process.}, vol. 60, no. 12, pp. 6202--6216, Dec. 2012.
  
\bibitem{li2015sufficient}
B.~Li, Y.~Shen, Z.~Wu, and J.~Li,
\newblock ``Sufficient conditions for generalized orthogonal matching pursuit
  in noisy case,''
\newblock {\em Signal Process.}, vol. 108, pp. 111--123, Mar. 2015.

\bibitem{mo2012remarks}
Q.~Mo and Y.~Shen,
\newblock ``A remark on the restricted isometry property in orthogonal matching
  pursuit algorithm,''
\newblock {\em IEEE Trans. Inf. Theory}, vol. 58, no. 6, pp. 3654--3656,
  Jun. 2012. 

\bibitem{chang2014improved}
L.~Chang and J.~Wu,
\newblock ``An improved \text{RIP}-based performance guarantee for sparse
  signal recovery via orthogonal matching pursuit,''
\newblock {\em IEEE Trans. Inf. Theory}, vol. 60, no. 9, pp. 5702--5715,
  Sep. 2014.
  
\end{thebibliography}


\end{document}